\documentclass[graybox]{svmult}

\usepackage{mathptmx}       
\usepackage{helvet}         
\usepackage{courier}        
\usepackage{type1cm}        
\usepackage{makeidx}         
\usepackage{graphicx}        
\usepackage{multicol}        
\usepackage[bottom]{footmisc}

\usepackage{caption}

\makeindex             

\begin{document}

\title*{Monitoring Jitter in Software Defined Networks}
\author{Jithin Kallukalam Sojan and K Haribabu}
\institute{Jithin Kallukalam Sojan \at Jithin Kallukalam Sojan, BITS Pilani, Rajasthan, India \email{f20170163@pilani.bits-pilani.ac.in}
\and K Haribabu \at K Haribabu, BITS Pilani, Rajasthan, India \email{khari@pilani.bits-pilani.ac.in}}
%
%
\maketitle

\abstract*{End-to-end jitter of a flow is an important metric that indicates the Quality of Service a user is experiencing, particularly for real-time applications such as video streaming, cloud gaming, and so on. Monitoring the jitter can help controllers make routing decisions for certain flows. This paper discusses methods in which we can estimate/follow important patterns in the end-to-end jitter of a flow. Two main approaches are taken, one using a Software Defined Network controller and the other using P4 programmable data-planes. Results from the simulations of each method are discussed.}

\abstract{End-to-end jitter of a flow is an important metric that indicates the Quality of Service a user is experiencing, particularly for real-time applications such as video streaming, cloud gaming, and so on. Monitoring the jitter can help controllers make routing decisions for certain flows. This paper discusses methods in which we can estimate/follow important patterns in the end-to-end jitter of a flow. Two main approaches are taken, one using a Software Defined Network controller and the other using P4 programmable data-planes. Results from the simulations of each method are discussed.}

\section{Introduction}
\label{sec:1}
\subsection{End-to-End jitter}
\label{subsec:1}
End-to-end jitter between hosts in a network has multiple definitions, the most common being the difference in time for consecutive packets to reach from one host to the other over a number of network nodes, i.e., the difference in one-way delays between end hosts for consecutive packets. Jitter is of concern to the Quality of Service realization in a network.

Even when packets are sent from a host at a uniform rate, it is possible that the packets are not received by an end host at a uniform rate (zero jitter). This behaviour is the result of various characteristics of the traffic at different network nodes. According to previous works \cite{SimpleFormulaFor}, it can be accounted for by the difference in the in-time and out-time of packets at each node.

\subsection{Motivation behind monitoring jitter}
\label{subsec:2}
Most applications that rely on continuous transmission of packets like video/audio streaming, video chats and cloud gaming \cite{sdnController} are affected by jitter. Jitter at the end host deteriorates the quality of the user experience.

In order to minimize the jitter by taking real time decisions within a network, such as changing the path of a flow, we need to estimate the end-to-end jitter experienced. Being able to closely estimate the jitter (or patterns in the jitter) at multiple hops away from the end-host is thus important.

\subsection{Approaches taken to monitor the jitter}
\label{subsec:3}
Our first attempt to monitor the jitter involves using a common controller (Floodlight) that polls for  statistics already collected by OVS switches, such as the count of packets that have passed through a switch (or a port of the switch) within a particular interval of time.

Later, we switch our approach to using only local information at the switch level. This involves using programmable data-planes (P4 programmable switches) to estimate the jitter at the data-plane itself. We can then send these estimates to a controller for further use. The switch can also be programmed to make real-time decisions based on the statistics it collects and calculates.

\section{Related Work}
\label{sec:2}
In \cite{passiveTCPStream}, RTT estimation at the mid-point of a TCP stream is discussed. Jitter is calculated as the absolute difference between the estimated RTT and the actual RTT calculated at that instant. Mean jitter is defined as the average of all the estimates for that particular stream. The authors give a direct formula to estimate the jitter.
\begin{equation}
Jitter_{t+1} = \frac{7}{8}.Jitter_{t} + \frac{1}{8}.Jitter_{sample}\;
\end{equation}
Compared to the focus of \cite{passiveTCPStream}, we direct our attention to the jitter experienced in the use of real time protocols such as UDP, as the objective is to monitor jitter that can affect the QoS a user experiences.

In \cite{AnAnalyticalModel}, the end-to-end jitter of a tagged stream in a tandem queuing network is discussed. The authors show various results, including the fact that the jitter decreases with increasing the load, and that the total jitter depends on the position of the congested nodes in the path. The authors assume a Poisson distribution for each stream, and mathematically derive the average jitter of an entire flow. Although it discusses estimating the jitter in a multi-node case, traffic with a Poisson distribution cannot be assumed for a modern network. As compared to a model that calculates the average jitter for an entire flow, we aim to monitor the jitter continuously.

In \cite{SimpleFormulaFor}, a strong mathematical formulation for the end-to-end jitter is discussed with the assumption of periodic traffic. The end-to-end jitter is given by the expected absolute value of the sum of inter-packet delay variations introduced by each node along the path. It introduces an exponential approximation for the steady state waiting time, which is used in the estimations of the jitter for both single and multi-node cases, but the jitter estimation is limited by the assumption on waiting times between two consecutive packets.

RFC 1889 \cite{rfcRTP} discusses a Real Time Protocol and some traffic characteristics related to it. The jitter at a packet arrival is calculated as the sum of $1/16$ times the jitter caused by the current and the previous packet and $15/16$ times the previous estimate of the jitter. We use this idea of exponential averaging to estimate the jitter in our P4 (programmable data-plane) implementation.

In \cite{delayAndJitterAnalysis}, the authors analyze the delay and jitter in networks that handle huge traffic volumes. Although the work is not on directly estimating the jitter, the authors give an analytical formulation for jitter with the assumption of Poisson traffic. It discusses the effect on end-to-end jitter as a result of changing network parameters such as increasing the network size, increasing the load, etc.

In \cite{sdnController}, the authors discuss latency and jitter in Cloud-Gaming, an example of an application where jitter estimation and actions based on the estimate are important. Although this paper does not directly talk about the estimation of jitter, it proposes a Load-Sharing algorithm that can be employed in an SDN controller to greatly decrease the jitter experienced in certain cases.

\section{Mathematical Formulation}
\label{sec:3}
\subsection{Nodes adding variation in the one-way delay}
\label{subsec:4}
Previous works \cite{SimpleFormulaFor} mention the idea that the end-to-end jitter can be modelled as the amount of variation each node in the network adds to the flow. Using SDN controllers we can collect real-time statistics such as the count of packets/bytes that have passed through a port for a particular flow. We collect the count of packets that have been transmitted by a port x of a switch S. Let this count of packets be a. Next, we collect the same statistic after t seconds. Let this count of packets be b. Thus, the rate at which packets were transmitted by port x of Switch S is: 
\begin{equation}
r = \frac{(b-a)}{t}\;
\end{equation}
Or, over the interval t, the average time of upload of a packet into the link at port x of Switch S was: 
\begin{equation}
t_{offload,x,S} = \frac{t}{(b-a)} \;
\end{equation}

Consider a network with two switches. Both S1 and S2 have in-ports x1 and x2 and out-ports y1 and y2. Host h1 is connected to S1 and host h2 is connected to S2, a flow from h1 to h2 is considered. Therefore the total time that a packet takes from h1 to h2 (one-way delay) is:
\begin{equation}
t_{1} = t_{h1,S1} + t_{load,x1,S1} + t_{offload,y1,S1} + t_{S1,S2} + t_{load,x2,S2} + t_{offload,y2,S2} + t_{S2,h2}\;
\end{equation}
Similarly, the one-way delay for the next packet will be:
\begin{equation}
t_{2} = t_{h1,S1} + t^{'}_{load,x1,S1} + t^{'}_{offload,y1,S1} + t_{S1,S2} + t^{'}_{load,x2,S2} + t^{'}_{offload,y2,S2} + t_{S2,h2}\;
\end{equation}
It is safe to assume that the time it takes for packets to travel across a link is constant. Thus, the only variable times in the above equations would be the load and offload times. The end-to-end jitter as a result of these two packets would be: 
\begin{equation}
t_{2}-t_{1} =  (\delta t_{load,x1,S1}) + (\delta t_{offload,y1,S1}) +(\delta t_{load,x2,S2}) + (\delta t_{offload,y2,S2})\;
\end{equation}
This equation is developed with the idea that we can model the jitter as the addition of variations in the one-way delay of a flow by the network nodes in the path. The difference here is that instead of just network nodes, we are considering each port (in-port and out-port) of each network node in the path.

Similar to the equation above, the current estimate of the delay value can be subtracted from the delay value that was calculated during the previous iteration. This would give us a statistic to monitor jitter, as we will see in Section \ref{sec:5}. This statistic can be easily calculated by a controller using the count of packets that have been transmitted or received at a particular port over an interval t.

\subsection{Deviation from the one-way delay}
\label{subsec:5}
As compared to a simple SDN controller setup, it is possible to calculate the actual one-way delay from the end-host to various network nodes in a P4 data-plane. If any node (P4 switch) between the sending host and the receiving host has collected the previous $n-1$ one-way delays that the packets took to reach that switch, the mean one-way delay can be calculated on the current packet arrival as: 
\begin{equation}
t_{mean} = \frac{\sum_{i=1}^{n} t_{i}}{n}\;
\end{equation}
As indicated by previous works \cite{passiveTCPStream}, the jitter can be considered as the standard deviation from the mean of the one-way delay. This standard deviation can be calculated as: 
\begin{equation}
\delta t = \sqrt{\frac{\sum_{i=1}^{n} (t_{i}-t_{mean})^{2}}{n-1}}\;
\end{equation}

Thus, we can calculate an estimate for the jitter using a data-plane programmable node that is able to retain the past $n-1$ one-way delays of packets that it received. At the same time, there are limitations to most P4 programmable switches that we need to account for in order to estimate the jitter.

Firstly, many P4 programmable switches cannot perform normal division. Division by a power of 2 is possible, emulated by bit-shifts to the right. Since the degree of freedom for the variance of a set is n-1 where n is the number of data-points, we consider n in our experiments to be a power of 2 plus 1, i.e., $ n = 2^{m}+1 $. Since the mean of the one-way delays also have to be calculated, we approximate it to the mean of the last n-1 packets.

Secondly, P4 programmable switches cannot be used to calculate square roots. Thus, in our experiments, we compare the squares of the actual value of the end-to-end jitter with the variance of the one-way delay, i.e., $(\delta t)^{2}$.

Finally, squaring a number is a computationally expensive task that should be avoided in the P4 programmable switch if possible. We thus attempt to use a deviation from the mean that is based on Manhattan distance, rather than Euclidean distance (standard deviation). Manhattan distance is given by the absolute value of the difference in two vectors.

The deviation based on the Manhattan distance will be calculated as:
\begin{equation}
\delta t = \frac{\sum_{i=1}^{n}|t_{i}-t_{mean}|}{n-1}\;
\end{equation}

\section{SDN Controller Implementation and Results}
\label{sec:4}
\subsection{Implementation of the Jitter Estimator}
\label{subsec:6}
\begin{figure}[htbp]
	\centering
		\includegraphics[width=80mm,scale=0.65]{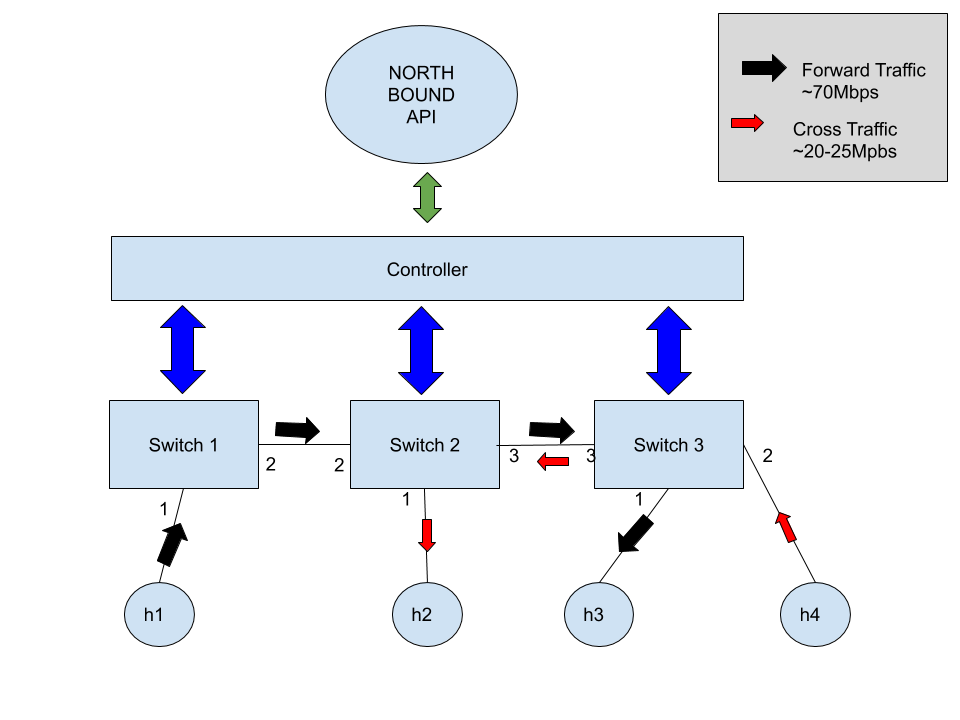}
	\caption[Topology used in the SDN approach]{Topology used in SDN approach (100 Mbps links between switches)}
	\label{fig:Topology used in SDN approach}
\end{figure}

Following the first mathematical formulation, we calculate the average time that it takes for a packet of a flow to offload into a link from a port and to load from a link into a port. We thus consider the average time it takes for a packet to offload from port2 of S1, port3 of S2 and port1 of S3 (refer Fig. \ref{fig:Topology used in SDN approach}). We also consider the average time it takes for a packet to load into port1 of S1, port2 of S2 and port3 of S3. We subtract the sum of these average times from the sum of the average times corresponding to the same ports from the previous iteration of statistics collection.

The experiments in this section were run on OVS switches in a Mininet network. A Floodlight SDN controller was used to communicate with the switches. The links between the three OVS switches were configured at a 100 Mbps each.

\subsection{Single Flow Runs}
\label{subsec:12}
\begin{figure}[htbp]
    \centering
		\includegraphics[width=90mm,scale=0.7]{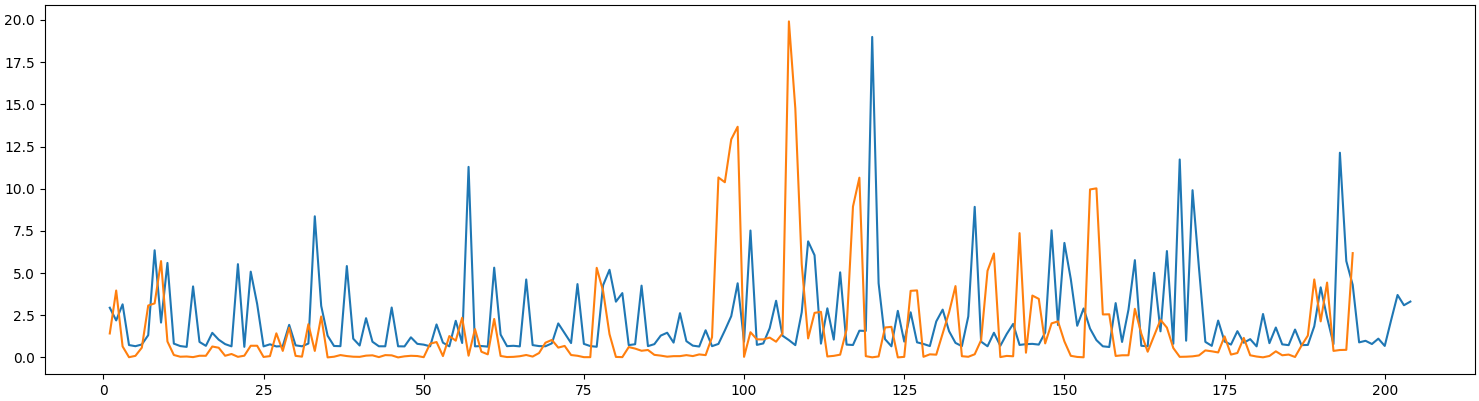}
	   \caption[Single Flow 90Mbps]{Single Flow 90 Mbps}
	\label{fig:Single Flow 90Mbps}
\end{figure}

First, a 90 Mbps flow was run from h1 to h3 for 200 seconds, with statistics polled every 1 second. On plotting the the end-to-end jitter actually experienced along with the results of the estimate against time, we get Fig. \ref{fig:Single Flow 90Mbps} (orange represents the estimate while blue represents the actual jitter).

It is impossible to synchronize these two time series using an SDN controller. This is because the estimates of the jitter are completely time-independent of the jitter calculation at the host. Thus, one can only approximately place the two time series together for visual comparison. The measure of similarity between the time series is done using a distance measure called Dynamic Time Warping (DTW). It is a measure of the minimum distance between the two time series, even when they are not synchronized in time. For the above mentioned run, the DTW distance between the estimate and the actual jitter experienced at the host is 327.206 units.

Although the base of the estimate seems to be lower than that of the actual end-to-end jitter, the peaks in the jitter, which is of concern to us, are almost of the same height. Also, the number of peaks are almost the same in the estimate as in the actual jitter, although the peaks have been depicted a bit forward in time, and sometimes at different heights. The ability to capture the surges in jitter is extremely helpful as the controller can use this information to make real time decisions in order to reduce the jitter.

Note that for all plots in this section, the above justifications hold. Thus we will be evaluating the runs based on the similarity of the peaks in the estimate and the actual jitter, and also based on the DTW distance between the two time series.

\begin{figure}[htbp]
	\centering
		\includegraphics[width=90mm,scale=0.7]{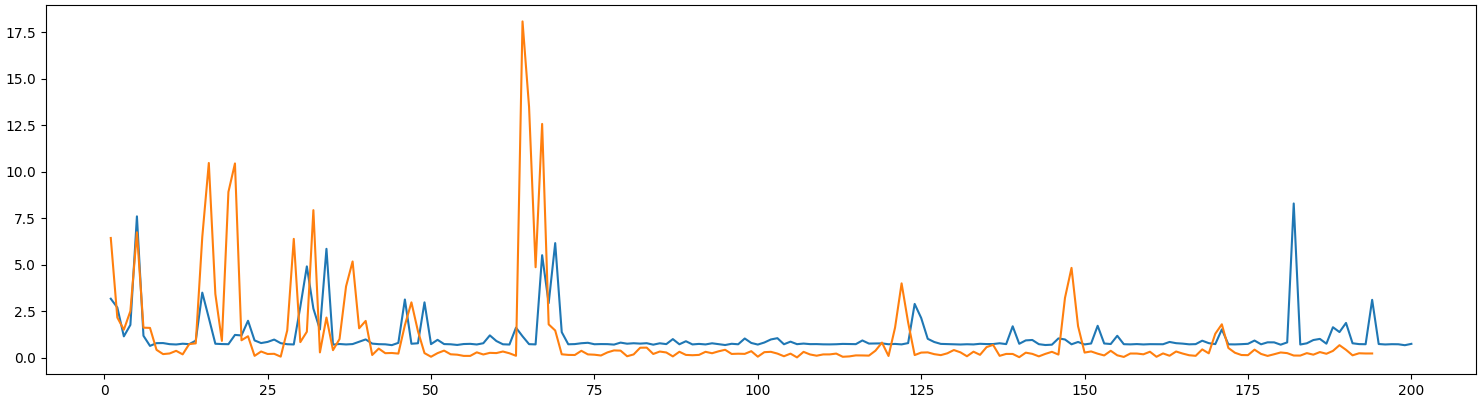}
	\caption[Single Flow 50Mbps]{Single Flow 50 Mbps}
	\label{fig:Single Flow 50Mbps}
\end{figure}

Next, we run a single flow of 50 Mbps between h1 and h3. The results are shown in Fig. \ref{fig:Single Flow 50Mbps}.

Although the peaks of the estimate are higher than that of the actual jitter, most of the surges in the estimate seem to have a corresponding surge in the actual jitter. The DTW distance between the estimate and the actual jitter turns out to be 201.516 units. This indicates that when the bandwidth utilization is decreased, the estimates seem to be getting better.

\begin{figure}[htbp]
	\centering
		\includegraphics[width=90mm,scale=0.7]{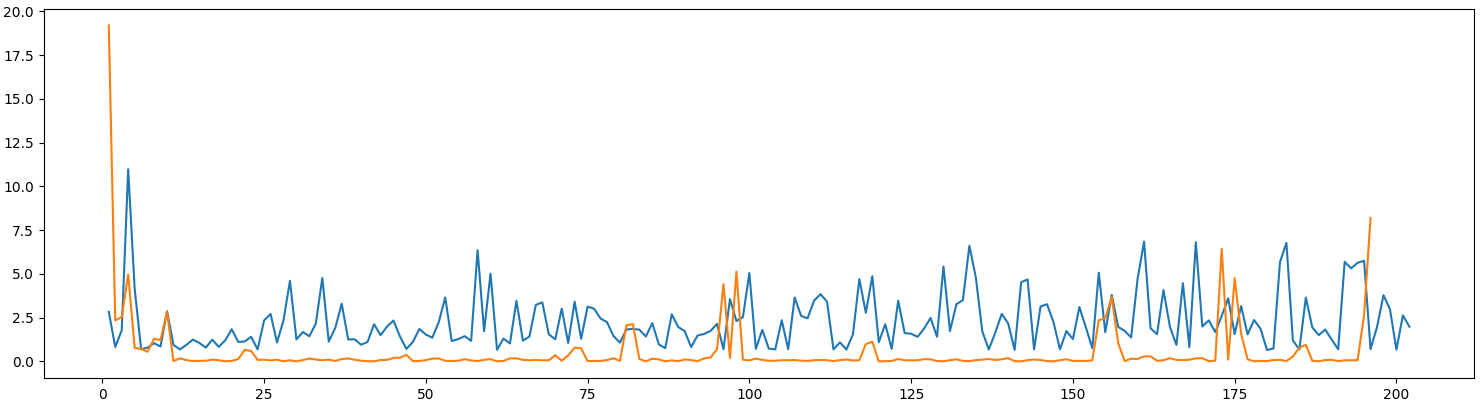}
	\caption[Single Flow 100Mbps]{Single Flow 100 Mbps}
	\label{fig:Single Flow 100Mbps}
\end{figure}

Next, we run a single flow of 100 Mbps between h1 and h3. The results are shown in Fig. \ref{fig:Single Flow 100Mbps}.

The DTW distance between the estimate and the actual jitter in this run is 345.85 units. Since a 100Mbps flow is not healthy over a 100Mps link, a number of packets are lost. This again indicates that on increasing the bandwidth utilization of the flow, the estimate tends to become worse.

\subsection{Increasing the period of Statistics collection}
\label{subsec:7}
The period of statistics collection is generally increased to counter an inherent problem of statistics collection at smaller time intervals, i.e., the likelihood of spurious values for the traffic statistics collected and the resultant calculations. On increasing the time interval over which the next statistic is collected, we smoothen out the values of the statistics in question.

\begin{figure}[htbp]
	\centering
		\includegraphics[width=90mm,scale=0.7]{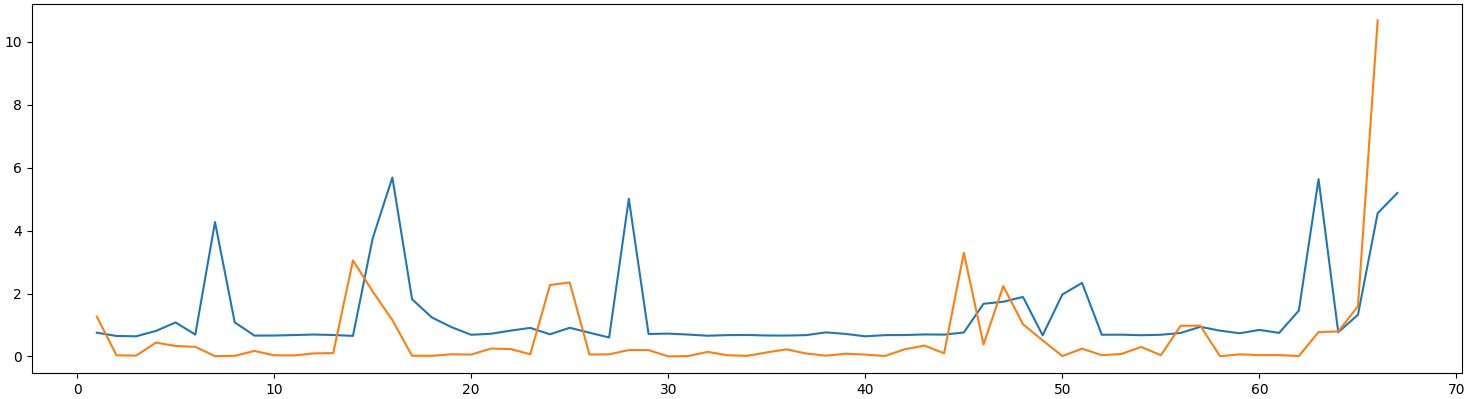}
	\caption[90Mbps Single Flow - Interval 3s]{Single Flow 90 Mbps- Interval 3s}
	\label{fig:90Mbps Single Flow - Interval 3s}
\end{figure}

We run a 90 Mbps flow from h1 to h3 for 200 seconds, with statistics from the ports of interest collected every 3 seconds. This is compared with the end-to-end jitter experienced by the flow over periods of 3 seconds as shown in Fig. \ref{fig:90Mbps Single Flow - Interval 3s}.

The DTW distance between the two time series is 61.66. Each surge seems to have a corresponding surge in the actual jitter experienced. Since the estimate is calculated over a different period of statistics collection from the other runs, a direct comparison between the DTW distances cannot be made (the DTW distance per point is much lesser in this run than the previous ones).

\section{P4 Implementation and Results}
\label{sec:5}
\subsection{Implementation of the Jitter Estimator}
\label{subsec:8}
This approach aims at estimating the end-to-end jitter within a programmable data-plane setup in a computationally inexpensive manner. As discussed in Section \ref{sec:3}, we save one-way delays from the source host at respective switches for $2^{n}+1$ packets. These delays are used to estimate the jitter according to the formulae given in Section \ref{sec:3}. Since the one-way delays are saved until $2^{n}+1$ packets arrive, the jitter estimator and thus the extra computation is used only once in these intervals.

The topology used for the experiments below is similar to that in the previous section. Three P4 programmable BmV2 switches S1, S2 and S3 are linearly connected in a Mininet network. We do not need to use P4 run-time as these experiments are run on static flows. A host h1 is connected to S1 and a host h2 is connected to S3. All the runs consists of a flow from h1 to h2. In order to collect the timestamps of when the packets are sent, an extra header dedicated to timestamps is used.

\subsection{Euclidean Distance estimate}
\label{subsec:9}
\begin{figure}[htbp]
	\centering
		\includegraphics[width=90mm,scale=0.7]{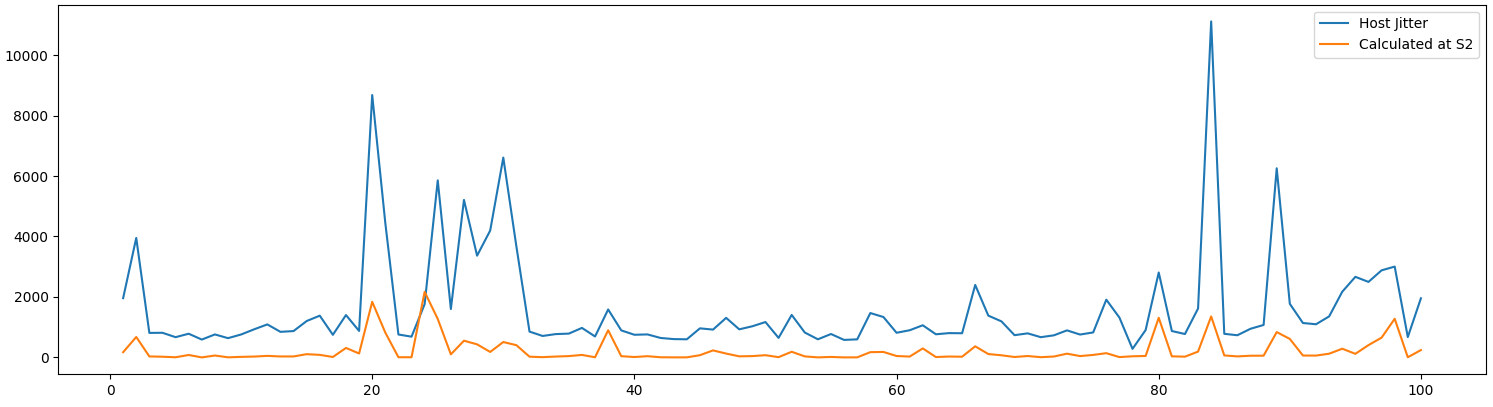}
	\caption[Euclidean Distance Variance at S2 (17 packets)]{Euclidean Distance Variance at S2 (17 packets)}
	\label{fig:Euclidean Distance Variance-17 packets-S2}
\end{figure}

As explained in Section \ref{sec:3}, the variance in mean one-way delay is an estimate of the square of the average jitter experienced over $2^{n}+1$ packets. Fig. \ref{fig:Euclidean Distance Variance-17 packets-S2} shows a comparison of the estimates from switch S2 and the actual jitter.

Unlike in the SDN approach, there is no discrepancy regarding where the two series should be placed with respect to each other. This helps with the visual comparison of the two series. Although the base of the estimate is lower than that of the actual jitter, the peaks of the estimate fall right in place with the peaks of the actual end-to-end jitter experienced in the network. The distance between the two time series shown in the plot Fig. \ref{fig:Euclidean Distance Variance-17 packets-S2} is 125805 units. This value is large because we are comparing squares of the value of jitters.

\begin{figure}[htbp]
	\centering
		\includegraphics[width=90mm,scale=0.7]{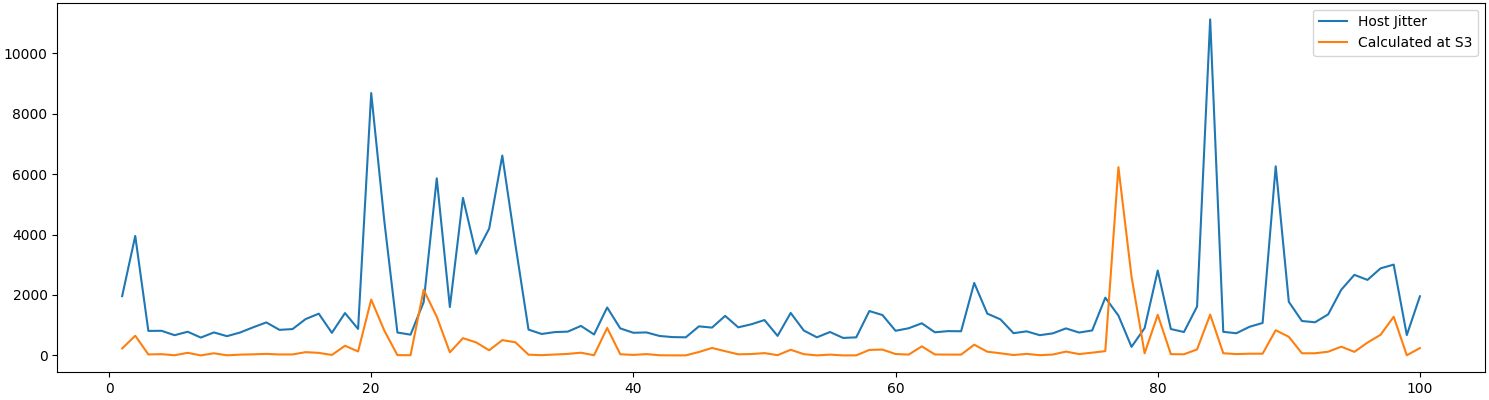}
	\caption[Euclidean Distance Variance at S3 (17 packets)]{Euclidean Distance Variance at S3 (17 packets)}
	\label{fig:Euclidean Distance Variance-17 packets-S3}
\end{figure}

For the same run as the one above, we estimate the jitter at Switch S3 and compare it with the square of the actual end-to-end jitters. Fig. \ref{fig:Euclidean Distance Variance-17 packets-S3} shows the results.

The results are almost the same as in the previous case with the peaks at the same places, except for a few random outliers. The distance between the estimate and square of the actual jitters is better than that of S2, at 123340.37 units. This indicates that as the number of hops to the receiving host decreases, the jitter estimates tend to get closer to the actual jitter.

\begin{figure}[htbp]
	\centering
		\includegraphics[width=90mm,scale=0.7]{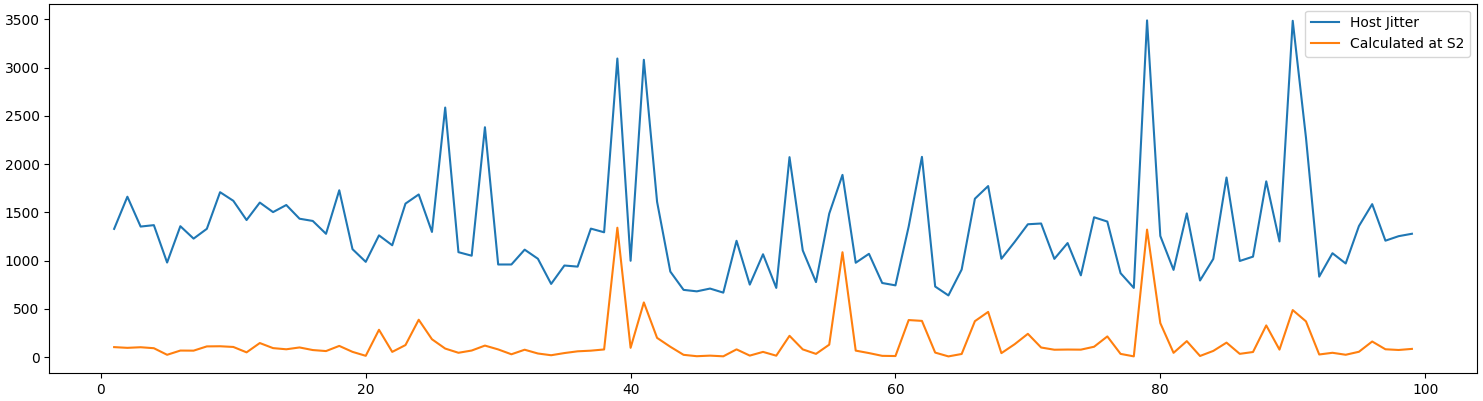}
	\caption[Euclidean Distance Variance at S2 (33 packets)]{Euclidean Distance Variance at S2 (33 packets)}
	\label{fig:Euclidean Distance Variance-33 packets-S2}
\end{figure}

Next we run a Euclidean Distance estimate for the square of the jitters, where the interval of estimation is at 33 packets. Fig. \ref{fig:Euclidean Distance Variance-33 packets-S2} shows the results.

As in the previous case, there seems to be a one is to one correspondence of the peaks. The distance between the two series is 111154.71 units, which is better than the estimates of both S2 and S3 in the 17 packets case. This indicates that as we increase the interval of jitter estimation, the estimate tends to get closer to the actual jitter.

\subsection{Manhattan Distance estimate}
\label{subsec:10}
\begin{figure}[htbp]
	\centering
		\includegraphics[width=90mm,scale=0.7]{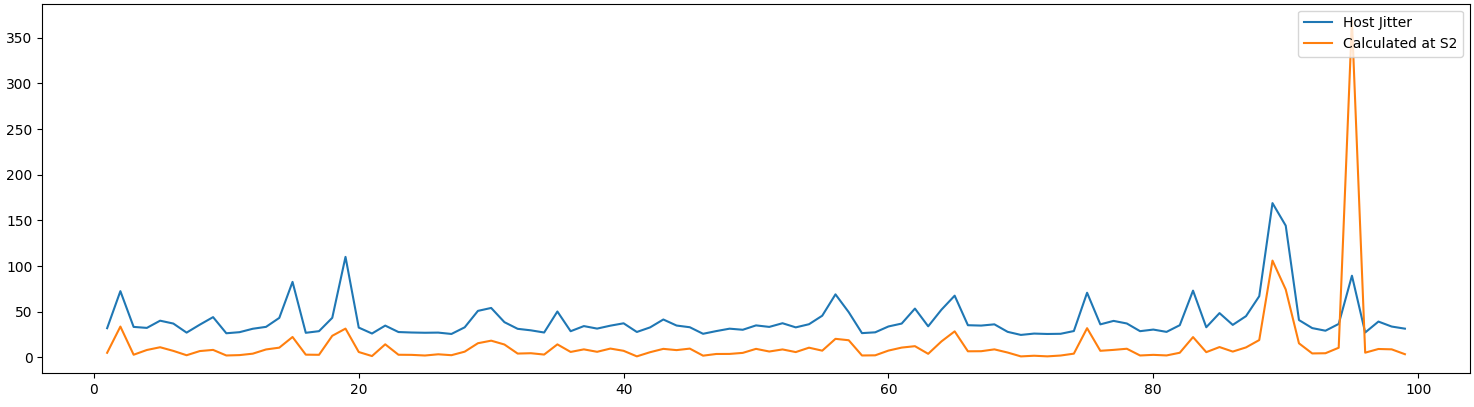}
	\caption[Manhattan Distance Estimate at S2 (17 packets)]{Manhattan Distance Estimate at S2 (17 packets)}
	\label{fig:Manhattan Distance Estimate-17 packets-S2}
\end{figure}

We estimate the jitter at switch S2 using a Manhattan distance estimate (ref. Section \ref{sec:3}) and at an interval of 17 packet arrivals. Fig. \ref{fig:Manhattan Distance Estimate-17 packets-S2} shows a comparison between the estimate and the actual end-to-end jitter.

The base of the actual jitter time series and the estimated jitter is much closer than in the Euclidean distance estimate (since the estimate is not on the squared jitter). There is a one-is-to-one correspondence between the peaks, as in the previous cases. The distance between the two time series is 3230.63 units. For the estimate at switch S3 in the same run, the distance is lower, at 3209.3 units.

\begin{figure}[htbp]
	\centering
		\includegraphics[width=90mm,scale=0.7]{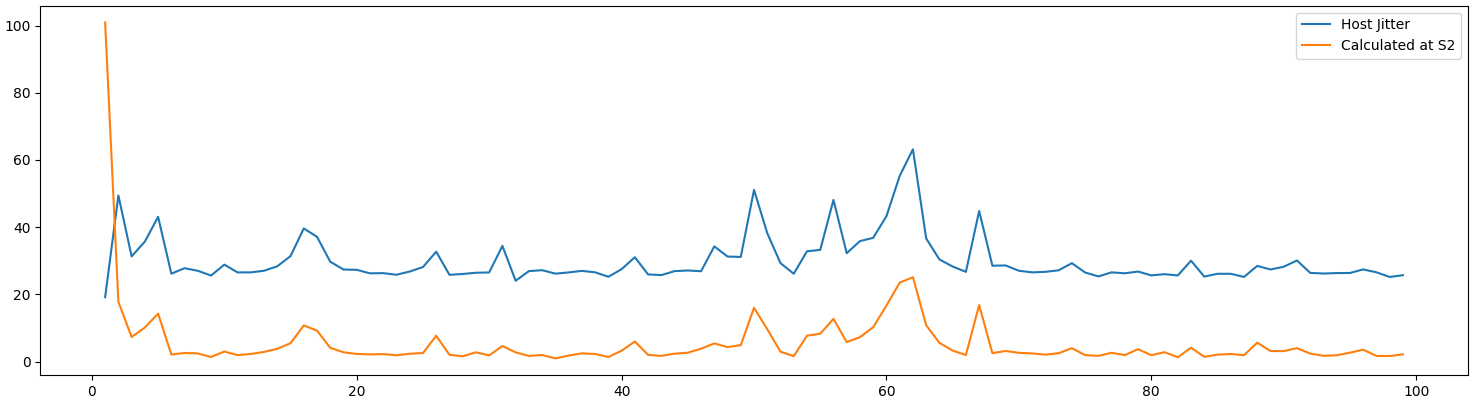}
	\caption[Manhattan Distance Estimate at S2 (33 packets)]{Manhattan Distance Estimate at S2 (33 packets)}
	\label{fig:Manhattan Distance Estimate-33 packets-S2}
\end{figure}

We also estimate the jitter at switch S2 at an interval of 33 packet arrivals, using a Manhattan distance estimate. Fig. \ref{fig:Manhattan Distance Estimate-33 packets-S2} shows a comparison between the estimate and the actual end-to-end jitter.

The distance between the two time series is 2558.14 units. This is lower than the distance which results from a 17 packet interval of estimation. Also, as expected, the distance of the estimate at S3 from the actual jitter is slightly lower at 2535.71.

\section{Conclusions}
\label{sec:6}
Modelling the end-to-end jitter experienced by a flow as the sum of variations in the one-way delay added by each each port in the path gives a statistic that lets the SDN controller monitor the jitter. It indicates surges in the jitter, which is the major motivation behind jitter estimation. Also, from the results of the simulations, it can be concluded that with a decrease in the link utilization of the flow, the jitter estimate gets closer to the actual jitter. Increasing the period of statistics collection results in less spurious values in the estimated jitter.

Modelling the end-to-end jitter as the standard deviation of one-way-delays between end-hosts is especially useful in programmable data-planes where the count of packets that have passed through and their timestamps can be stored. On increasing the number of packets over which the standard deviation is calculated, or on decreasing the hops to the receiving host, the estimate gets closer to the actual jitter experienced. Also, using Manhattan distance gives a decent estimate of the jitter with the added benefit of its computational lightness compared to the Euclidean distance estimate.

\end{document}